 \documentclass[twocolumn,showpacs,prl]{revtex4}

\usepackage{graphicx}% Include figure files
\usepackage{dcolumn}% Align table columns on decimal point
\usepackage{bm}% bold math 
\begin{document}

\title{Energetics of nitrogen incorporation reaction in SiO$_2$}
\author{Walter Orellana} 
\email{wmomunoz@if.usp.br}
\affiliation{Instituto de F\'{\i}sica, Universidade de S\~ao Paulo, 
C.P. 66318, 05315-970, S\~ao Paulo, SP, Brazil}

%date\today

\begin{abstract} 
We study using first-principles calculations the energetics, structural 
and electronic properties of nitrogen incorporation in SiO$_2$. 
We consider NO, NH, N$_2$ and atomic N as the nitriding species interacting 
with a Si-Si bond of an otherwise perfect SiO$_2$ network in order to 
simulate the nitrogen incorporation near Si/SiO$_2$ interface regions. 
We find that all the species react with the Si-Si bond forming bridge 
structures with the Si atoms without dissociating, where NH and atomic N 
form the most stable structures. 
Concerning the electronic properties, our results show that
incorporated NH is the only structure which does not introduce trapping
center at the interface. The structures involving NO and atomic N are acceptors,
whereas that involving N$_2$ may be either a donor or an acceptor.
The hydrogen passivation of the electrically active centers is also discussed.

\end{abstract} 
\pacs{61.72.Ji, 71.55.Ht, 71.15.Mb}

\maketitle
% \newpage

Oxynitride films have been extensively studied in the past years since
they improve the reliability of metal-oxide-semiconductors gate insulators. 
The main benefits of incorporating nitrogen into ultrathin SiO$_2$ films 
are the reduction of gate leakage currents and the resistance to boron 
penetration\cite{green}. 
The growth of ultrathin oxynitride films strongly depends on the reactant 
species ({\it e.g.}, N$_2$O, NO, NH$_3$, N$_2$) and the technique used. 
Nitrogen can be incorporated into SiO$_2$ using either thermal oxidation and
annealing or chemical and physical deposition methods. 
Thermal nitridation of SiO$_2$ in NO and N$_2$O generally 
results in a relatively low N concentration at the near-interface 
(Si/SiO$_2$) region\cite{ellis,baumvol}.
The N incorporation is commonly associated to the reaction 
of NO molecule with Si-Si bonds at the interface, after diffusing 
through the oxide\cite{lu}. On the other hand, N incorporation via 
annealing in NH$_3$ is responsible for a relatively high N concentrations 
into the films\cite{baumvol}. This method provides both near-interface and 
near-surface nitridation, suggesting different nitriding species derived 
from NH$_3$. 
More recently, higher N concentrations and controlled distributions have
been attained by plasma assisted methods, typically using ions and radicals 
derived from N$_2$ and NH$_3$ as nitrogen sources\cite{watanabe}. 
Although the control of both the density and the distribution of 
nitrogen into SiO$_2$ have been achieved at few-layer level, less is known
about the early stages of the N-incorporation reactions at atomic level.

In this work the energetics and structural properties of near-interface
nitrogen incorporation are studied from first-principles total-energy 
calculations. We have considered the N$_2$, NH and NO molecules as well 
as atomic N as the precursor species, reacting with the Si-Si bonds in 
SiO$_2$ in order to simulate suboxide (Si$^{3+}$) or near-interface N 
incorporation.
Our calculations were performed in the framework of the density functional 
theory, using a basis set of numerical atomic orbitals as 
implemented in the SIESTA code\cite{siesta}.
We have used a split-valence double-$\zeta$ basis set plus the polarization 
functions as well as standard norm-conserving pseudopotentials\cite{pseudo}.
For the exchange-correlation potential we adopt the generalized gradient 
approximation\cite{pbe}. We used a 72-atom $\alpha$-quartz supercell and the 
$\Gamma$ point for the Brillouin zone sampling. 
The positions of all the atoms in the supercell were relaxed until all the 
force components were smaller than 0.05 eV/$\text{\AA}$.
We also consider neutral and singly charged species, where the neutrality 
of the cell is always maintained by introducing a compensating background 
charge. Spin-polarization effects, which are important for the correct
description of atomic and molecular reaction processes in SiO$_2$\cite{prl2-us},
are included throughout the calculation. 

\begin{figure}
\includegraphics[width=7.0cm]{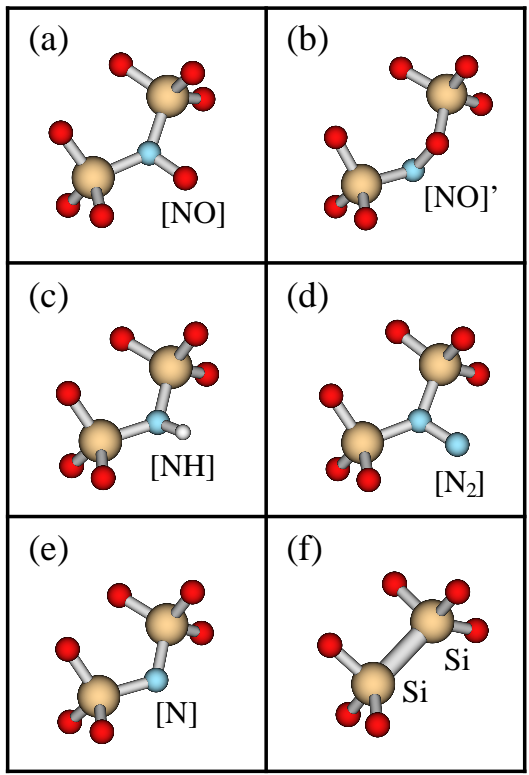}
\caption{\label{f1} Local equilibrium geometries for neutral nitriding 
species after reacting with Si-Si bond in SiO$_2$.
(a) the most stable geometry for the incorporated NO.
(b) the second stable geometry for the incorporated NO.
(c), (d), and (e) are the stable geometries for the
incorporated NH, N$_2$ and atomic N, respectively.
(f) the Si-Si bond or the suboxide region (Si$^{3+}$) in SiO$_2$.}
\end{figure}

We study the chemical reactions occurring when the nitriding species
NO, NH, N$_2$ and atomic N approach the Si-Si bond in the otherwise
perfect SiO$_2$ network. 
The Si-Si bond in SiO$_2$ is formed when an O atom is removed from 
the network characterizing an oxygen vacancy. This local geometry is 
close related to those found near the Si/SiO$_2$ interface\cite{prl2-us}.
Hereafter, the incorporated species will be called [NO], 
[N$_2$], [NH] and [N]. For these reactions we only consider neutral 
species in their ground-state spin configurations.
Initially, we study the structural properties of the nitriding species
inside the largest interstitial site of a perfect SiO$_2$, exploring 
possible reactions that they may undergo with the network.
Our results show that none of the species considered in this work
react with the network, remaining at the interstitial sites. This suggests 
that neutral species would be diffusing species in SiO$_2$. However, when 
we put the species close to the Si-Si bond, they are quickly incorporated 
into the network forming the stable structures shown in Fig.~\ref{f1}. We
now describe in details our results for each incorporated specie.
%%%%%%%%%%%%%%%%%%%%%%%%%%%%%%%%%%%%%%%%%%%%%%%%%%%%%%%%%%%%%%%%%%%
%%%%%%%%%%%%%%%%%%%%%%%%%%%%%%%%%%%%%%%%%%%%%%%%%%%%%%%%%%%%%%%%%%%

The NO molecule shows two stable structures after reacting with the Si-Si
bond. In the lowest-energy structure [NO], the N atom is threefold coordinated 
bonding with two Si atom, keeping the bond with the O atom, as shown in 
Fig.~\ref{f1}(a). 
Here, the Si-N and N-O bond lengths are 1.78 and 1.32~\AA, respectively.
The binding energy of NO, calculated as the difference in energy between 
the interstitial and incorporated configurations, is found to be 3.8~eV.
Whereas, the binding energy of the O atom in the [NO] structure is 4.2~eV.
Thus, it would be easier to remove 
the entire NO molecules than the single O atom. Fig.~\ref{f1}(b) shows
the second stable geometry [NO]'. Here, NO is incorporated into the Si-Si bond 
forming a structure similar to the peroxyl bridge of 
oxygens\cite{hamann,prl2-us}. 
This structure is 0.65~eV higher in energy than the most stable one. The 
Si-N, N-O, and O-Si bond lengths are 1.77, 1.40, and 1.71~\AA, 
respectively.  The energy barrier for the NO molecule to change from the
metastable structure [NO]' to the most stable one [NO] is estimated to be 
0.6~eV. The lowest-energy [NO] structure is an electrically active center, 
exhibiting a half-occupied energy level at 1.9~eV above the density-functional 
valence-band maximum (VBM) of $\alpha$-quartz.
The [NO] dangling bond may be passivated by capturing an H atom as 
recently suggested\cite{jeong}. We find that the reaction involving the 
capture of an H atom from an interstitial H$_2$ molecule is endothermic with 
an energy cost of 0.78~eV. 
However, the [NO] passivation may occur by the capture of an
interstitial H atom. In this case, the reaction is highly exothermic with an 
energy gain of 4.1~eV.

Figure~\ref{f1}(c) shows the equilibrium geometry for the NH after reacting 
with the Si-Si bond in SiO$_2$. 
We see that the N atom binds to both Si atoms forming the Si-N-Si structure, 
keeping the bond with the H atom. The Si-N and N-H bond lengths 
are 1.72 and 1.03~\AA, respectively. The binding energy of the NH molecule is 
found to be 6.8~eV which indicates that this radical would be one of 
the most stable species for interface nitridation. We also found that the 
H atom is strongly bound to the N atom in [NH], having a binding energy of 
4.6~eV.

The equilibrium geometry for the incorporated N$_2$ is shown in 
Fig.~\ref{f1}(d). The N$_2$ binding energy is found to be 1.2~eV, 
relatively small as compared with the other nitriding species. This suggests 
that N$_2$ may be easily removed from the interface escaping from the 
SiO$_2$ film. Additionally, the [N$_2$] structure introduces an empty midgap 
level at 2.7~eV above the VBM of $\alpha$-quartz.

For the N atom we also find a very stable structure after reacting with
the Si-Si bond [Fig.~\ref{f1}(f)]. The binding energy of the N atom 
is found to be 6.5~eV. [N] shows a half-occupied 
energy level at 0.2~eV above the VBM of $\alpha$-quartz. This dangling
bond may be passivated by capturing an H atom from an interstitial H$_2$ 
molecule which dissociates after the contact with the [N] structure.
We find that this reaction is exothermic with an energy gain of 0.82~eV, 
where the resulting equilibrium structure is similar to the [NH] structure 
[see Fig.~\ref{f1}(c)] plus an interstitial H atom. 
If the [N] passivation occurs by capturing an interstitial 
H atom, the reaction is highly exothermic with an energy gain of 5.0~eV.

\begin{figure}
\includegraphics[width=8.5cm]{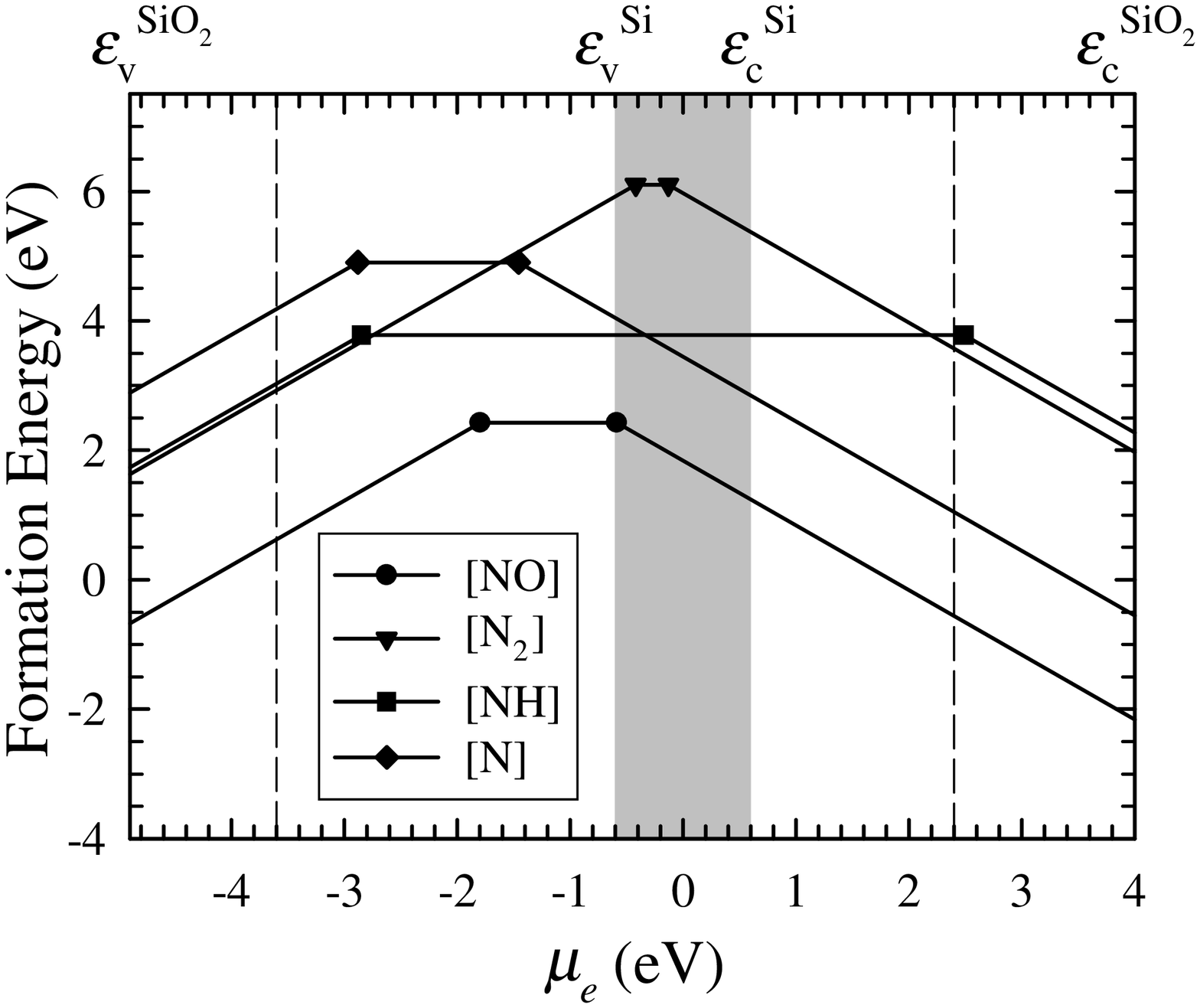}
\caption{\label{f2} Formation energies of the nitriding species after 
reacting with the Si-Si bond in SiO$_2$, as a function of the electron 
chemical potential ($\mu_e$) in the energy gap of SiO$_2$. 
$\varepsilon_{\rm v}^{\rm SiO_2}$ [$\varepsilon_{\rm c}^{\rm SiO_2}$] 
is the experimental valence-band maximum (VBM) [conduction-band minimum 
(CBM)] of $\alpha$-quartz. The dashed lines indicate the density-functional 
VBM and CBM and the shaded area represents the experimental band gap of 
bulk Si. The Si midgap is chosen as the zero energy of $\mu_e$.}
\end{figure}

The relative stability of incorporated nitriding species in SiO$_2$ in
thermodynamic equilibrium is obtained by comparing their formation 
energies. As we are simulating the nitrogen incorporation at suboxide 
regions, i.e., close to the Si/SiO$_2$ interface, we consider the three 
most relevant charge states in order to address possible charge transfer 
between the incorporated species and the doped Si substrate. 
Therefore, the formation energies ($E_{f}$) are calculated as a
function of the chemical potential of the nitriding species 
($\mu_{\rm X}$) and the electron chemical potential ($\mu_e$) as
\begin{equation}
E_{f}({\rm [X]}^{q},\mu_e)=E_{t}({\rm [X]}^{q})-E_{t}({\rm SiO_{2}})+
\mu_{\rm O}-\mu_{\rm X}+q\,\mu_{e},
\end{equation}
where $E_{t}({\rm [X]}^{q})$ is the total energy of the incorporated specie
X in the charge state $q$, $E_{t}$(SiO$_{2}$) is the total energy of 
perfect $\alpha$-quartz and $\mu_{\rm O}$ the chemical potential of 
oxygen. $\mu_{\rm X}$ and $\mu_{\rm O}$ are fixed at the values of their 
stable gas phases.
Our results for the formation energies of [NO], [NH], [N$_2$], and [N] are 
shown in Fig.~\ref{f2}. 
%%%%%%%%%%%%%%%%%%%%%%%%%%%%%%%%%%%%%%%%%%%%%%%%%%%%%%%%%%%%%%%%%%%
%%%%%%%%%%%%%%%%%%%%%%%%%%%%%%%%%%%%%%%%%%%%%%%%%%%%%%%%%%%%%%%%%%%
In the figure, the symbols indicate the transition states or the 
thermodynamics levels, whereas the slopes of the lines (positive, 
zero and negative) indicate the corresponding charge states. 
Following previous calculations\cite{blochl,jeong}, the transition 
states were aligned with respect to the experimental band-gap edges of
$\alpha$-quartz, by matching the calculated ($+/-$) transition state 
of the interstitial H atom in $\alpha$-quartz with its measured value 
of 0.2~eV above the Si midgap\cite{stahlbush,cartier}, and by using 
the measured valence-band offset of the Si/SiO$_2$ interface of 
4.3~eV\cite{himpsel}.
In Fig.~\ref{f2} we also depict the experimental band gap of bulk Si where
the Si midgap position is chosen to be the zero energy of $\mu_e$. 

As we are computing the energetics of the incorporated species close
to the Si/SiO$_2$ interface, the variation of $\mu_e$ must be considered
only in the energy range of bulk Si.
Fig.~\ref{f2} shows that negatively charged [NO] is the most stable structure 
at the interface. The ($0/-$) transition state is found at 4.3~eV above the 
experimental SiO$_2$ valence-band maximum 
($\varepsilon_{\rm v}^{\rm SiO_{2}}$). 
This suggests that [NO] might capture an electron from the Si substrate 
characterizing an acceptor center at the Si/SiO$_2$ interface. 
The same behavior is found for [N] which might be another acceptor center 
at the interface. For the case of [N$_2$] we find both donor and acceptor 
character, where the ($+/0$) and ($0/-$) transition states are localized at 
$\varepsilon_{\rm v}^{\rm SiO_{2}}$\,+\,4.5 and 
$\varepsilon_{\rm v}^{\rm SiO_{2}}$\,+\,4.8~eV, respectively.
The formation energy of [N$_2$] is approximately 2~eV higher in energy than
[NO], suggesting that [N$_2$] would be the less stable structure at the
interface. Finally, [NH] is found to be the only electrically inactive 
structure at the interface, having a relatively low formation energy. 
This suggests that the NH molecule would be the best choice among the 
considered nitriding species for interface nitridation.

In summary, we find that NH would be the most stable species for interface 
nitridation having a binding energy as high as 6.8~eV, where the N atom is 
typically threefold coordinated. NO would also be a very stable species at 
the interface having a binding energy of 3.8~eV. 
Our results agree well with the finding that NO diffuses through 
the oxide and reacts at the near-interface region where N is bound to two Si 
atoms and one O atom\cite{lu}. However, NO does not dissociate spontaneously
after reacting with a Si-Si bond. N$_2$ may also be incorporated at the 
interface, forming a stable structure with the Si-Si bond. However, it has 
a relatively low binding energy of 1.2~eV. This suggests that N$_2$ may be 
easily removed from the interface.
Atomic N does not react spontaneously with the SiO$_2$ network as commonly 
believed, suggesting that it would be a diffusing species in SiO$_2$. 
The binding energy of the incorporated N is 6.5~eV, resulting in a very 
stable species for interface nitridation. 
Concerning the electrical properties, our results show that NH is
the only species that does not introduce a trapping center after being
incorporated at the Si-Si bond. [NO] and [N] are acceptors whereas 
[N$_2$] may be a donor or an acceptor, depending on the Fermi level position. 
However, [N] and [NO] may be passivated by subsequent hydrogenation processes.

The author would like to thank Profs. A. Fazzio and Ant\^onio J. R. da 
Silva for fruitful discussions. This work was supported by FAPESP.

\end{document}